\documentclass[twocolumn]{openjournal}

\usepackage{xcolor}
\usepackage{textgreek}
\usepackage[utf8]{inputenc}
\usepackage[english]{babel}

\usepackage{hyperref}
\hypersetup{
    unicode, 
    colorlinks=true,
    linkcolor=linkcolor,
    citecolor=linkcolor,
    filecolor=linkcolor,
    urlcolor=linkcolor,
}
\usepackage{color,colortbl}
\definecolor{linkcolor}{rgb}{0.0,0.3,0.5}
\usepackage{tensind}
\tensordelimiter{?}
\DeclareGraphicsExtensions{.bmp,.png,.jpg,.pdf}
\usepackage{verbatim}
\usepackage[normalem]{ulem}
\usepackage{orcidlink}
\usepackage{soul}

\graphicspath{ {./figs/} }

\newcommand{\beq}{\begin{equation}}
\newcommand{\eeq}{\end{equation}}
\newcommand{\beqa}{\begin{equation}\begin{aligned}}
\newcommand{\eeqa}{\end{aligned}\end{equation}}
\newcommand{\bit}{\begin{itemize}}
\newcommand{\eit}{\end{itemize}}

\newcommand{\Msun}{M_\odot}

\begin{document}

\title{Cluster finding with outskirt stellar masses and percolation}

\author{Pablo Avalos}
\affiliation{Department of Physics and Astronomy, California State University, Northridge, 18111 Nordhoff Street, Northridge, CA 91330-8268, USA}
\affiliation{Department of Physics, University of California, Santa Cruz, Santa Cruz, CA 95064, USA}

\author{Conghao Zhou\,\orcidlink{0000-0002-2897-6326}}
\email{zhou.conghao@ucsc.edu}
\affiliation{Department of Physics, University of California, Santa Cruz, Santa Cruz, CA 95064, USA}
\affiliation{Santa Cruz Institute for Particle Physics, Santa Cruz, CA 95064, USA}

\author{Tesla Jeltema\,\orcidlink{0000-0001-6089-0365}}
\email{tesla@ucsc.edu}
\affiliation{Department of Physics, University of California, Santa Cruz, Santa Cruz, CA 95064, USA}
\affiliation{Santa Cruz Institute for Particle Physics, Santa Cruz, CA 95064, USA}

\author{Katya Leidig\,\orcidlink{0009-0003-8358-8320}}
\affiliation{Department of Astronomy, University of Maryland, College Park, MD 20742, USA}

\author{Shuo Xu\,\orcidlink{0000-0002-4460-0409}}
\affiliation{Department of Astronomy, Tsinghua University, Beijing 100084, China}

\author{Benedikt Diemer\,\orcidlink{0000-0001-9568-7287}}
\affiliation{Department of Astronomy, University of Maryland, College Park, MD 20742, USA}

\author{Song Huang\,\orcidlink{0000-0003-1385-7591}}
\affiliation{Department of Astronomy, Tsinghua University, Beijing 100084, China}

\author{Alexie Leauthaud\,\orcidlink{0000-0002-3677-3617}}
\affiliation{Department of Astronomy and Astrophysics, University of California, Santa Cruz, Santa Cruz, CA 95064, USA}

\begin{abstract}
The abundance of galaxy clusters is a powerful cosmological probe, but optical cluster cosmology is limited by selection systematics, in particular the projection effects that affect cluster finders based on galaxy populations such as the red sequence. The outer stellar mass ($M_\mathrm{out}$) of cluster central galaxies---e.g., the stellar mass in a 50--100 kpc annulus---offers an alternative selection that relies only on the central galaxy and is therefore largely free from projection effects. Its primary systematic is instead satellite contamination, since massive clusters can host more than one galaxy with high outer stellar mass. Using the IllustrisTNG300 simulation at $z=0.4$, we quantify this contamination and investigate a simple, proximity-based percolation method to mitigate it, in which galaxies with lower outer stellar mass lying within a given radius of a more massive galaxy are removed from the sample. We find that the satellite fraction defined by the friends-of-friends (FoF) algorithm is modest even without percolation ($\leq15\%$ for $M_\mathrm{out} > 10^{10}\,\mathrm{M}_{\odot}$ and $<10\%$ for $M_\mathrm{out} > 10^{11}\,\mathrm{M}_{\odot}$), and that percolation reduces it further, with the improvement increasing for percolation radii up to $3.0\,R_{200c}$. For a moderately high outer stellar mass cut ($\sim 4\times10^{10}\,\mathrm{M}_{\odot}$) and percolation radius ($\sim 2.0\,R_{200c}$), we recover a cluster sample that is both highly complete and pure for halo masses $\gtrsim 10^{14}\,\mathrm{M}_{\odot}$. These results indicate that outer stellar mass, combined with simple percolation, has the potential to provide a clean and readily calibratable selection of massive galaxy clusters.
\end{abstract}

\maketitle

\section{Introduction}

The abundance of clusters of galaxies and its evolution with redshift in surveys like the Legacy Survey of Space and Time (LSST) has the potential to strongly constrain the dark energy equation of state, and clusters offer a complementary probe that adds constraining power when combined with other methods \citep[e.g.][]{Weinberg2013, DES2020}.  One of the primary limitations in optical cluster cosmology stems from our understanding of and ability to calibrate the cluster selection function and its effect on mass calibration.  Methods of selecting clusters in photometric data based on their galaxy populations (e.g. the red sequence) inevitably suffer from projection effects due to line-of-sight structure \citep[e.g.][]{Lucey1983, Rozo2015b, Sohn2018, Costanzi2019, Sunayama20, Myles2021, Wu22, zhouForecastingConstraintsOptical2024}.  While progress is being made on understanding these effects \citep{wuOpticalSelectionBias2022, sunayamaOpticalClusterCosmology2024, zhouForecastingConstraintsOptical2024, myles25, salcedoCosmologicalConstraintsDark2025, Yang26}, it is worth considering alternative methods of selecting clusters that are relatively free from projection effects.

One promising potential method of selecting clusters is based on the large and extended stellar mass in cluster central galaxies \citep{huangOuterStellarMass2021, xhakajClusterCosmologyCluster2023, kwiecienImprovingGalaxyCluster2025, Zhou25, Xu25, leidigReachingEdgeII2025}.  The most massive galaxies in the universe lie preferentially in clusters, and central cluster galaxies host extended stellar distributions due to a history of mergers and accretion, which also builds the intracluster light (ICL) \citep{gonzalezIntraclusterLightNearby2005, zibettiIntergalacticStars0252005, huangIndividualStellarHaloes2018, wangStellarHaloIsolated2019, zhangDarkEnergySurvey2019, broughVeraRubinObservatory2020, klugePhotometricDissectionIntracluster2021, liReachingEdgeProbing2022, montesFaintLightGroups2022, zhangDarkEnergySurvey2024, Zhou25, leidigReachingEdgeII2025}.  Outer stellar mass ($M_\mathrm{out}$), defined as the stellar mass in an annulus from 50 to 100 kpc, has been shown to be an effective way of selecting clusters with low observable-to-halo-mass scatter comparable to leading red-sequence-based cluster finders (e.g. redMaPPer) \citep{huangOuterStellarMass2021, kwiecienImprovingGalaxyCluster2025}.  At the same time, this method, which uses only the central cluster galaxy, is free from projection effects induced by large-scale structure.  Therefore, outer stellar mass may offer a more easily calibratable cluster selection; it also shows promise as a method of selecting clusters down to comparatively lower masses than other selection techniques \citep{huangOuterStellarMass2021, kwiecienImprovingGalaxyCluster2025}.

However, outer stellar mass comes with its own selection systematics, in particular satellite contamination \citep{xhakajClusterCosmologyCluster2023, kwiecienImprovingGalaxyCluster2025}.  While satellite contamination is relatively low at the very highest stellar masses, massive clusters can host more than one galaxy with high stellar mass \citep{zehaviGalaxyClusteringCompleted2011, leauthaudNewConstraintsEvolution2012, reddickConnectionGalaxiesDark2013, zuMappingStellarContent2015, hoshinoLuminousRedGalaxies2015}.  This problem increases for the selection of lower-mass clusters, as there is an overlap in the stellar mass of centrals in group-size halos and satellite galaxies in clusters \citep{yangGalaxyGroupsSDSS2008, zuMappingStellarContent2015, tinkerSelfCalibratingHaloBasedGroup2021}. 

In this paper, we quantify satellite contamination in outer stellar mass-selected samples in simulations and investigate a simple, proximity-based percolation method for reducing this contamination.  We show that this percolation is effective at eliminating a significant fraction of satellite galaxies from the sample, and we find that for an appropriate outer stellar mass cut this yields complete and high-purity samples of the most massive halos in the simulation.  However, some amount of satellite contamination remains when pushing to lower outer stellar masses.

This paper is organized as follows.  In Section \ref{sec:sims}, we describe the IllustrisTNG300 simulation and the construction of the outer stellar mass catalog.  Section \ref{sec:method} presents the percolation method.  In Section \ref{sec:results}, we present the satellite fraction and the completeness and purity of the percolated samples as a function of the outer stellar mass cut and percolation radius.  In Section \ref{sec:discussion}, we discuss the impact of the friends-of-friends halo definition on the measured satellite contamination, and we conclude in Section \ref{sec:conclusions}.

\section{Simulations}
\label{sec:sims}
In this section, we describe the numerical simulation and the simulated cluster catalogs used in this work. We first introduce the IllustrisTNG simulations, and then we explain the production of the outer stellar mass catalogs. 

\subsection{Illustris-TNG300}

IllustrisTNG is a suite of magnetohydrodynamic simulations that simulate galaxies and galaxy clusters in a $\Lambda$CDM cosmology \citep{nelsonFirstResultsIllustrisTNG2018, pillepichFirstResultsIllustrisTNG2018, springelFirstResultsIllustrisTNG2018, nelsonIllustrisTNGSimulationsPublic2019}. IllustrisTNG employs the moving-mesh code \textsc{arepo} \citep{weinbergerAREPOPublicCode2020} to follow large-scale structure and galaxy assembly in three periodic volumes with side lengths of $\sim 50$, $\sim 100$, and $\sim 300$\, Mpc. The galaxy formation physics includes the modelling of radiative cooling and heating, star formation, stellar evolution,
galactic winds, and the formation, growth, and energetic feedback of supermassive black holes \citep{weinbergerSimulatingGalaxyFormation2017, pillepichSimulatingGalaxyFormation2018, naimanFirstResultsIllustrisTNG2018, nelsonFirstResultsIllustrisTNG2018}. The simulated observables agree well with the observational results \citep{pillepichFirstResultsIllustrisTNG2018, nelsonFirstResultsIllustrisTNG2018, springelFirstResultsIllustrisTNG2018, ardilaStellarWeakLensing2021, leidigReachingEdgeII2025}. It is known that the stellar mass from different TNG resolutions does not agree because of resolution effects \citep{pillepichFirstResultsIllustrisTNG2018}. We do not correct for this effect since the overall normalization of stellar mass does not impact the results of the current paper. In this work, we use TNG300-1, which is the highest mass resolution version of the largest box in the suite. The side length of the simulation is 302.6 comoving Mpc. The dark matter mass resolution and the baryon mass resolution are $4 \times 10^7 \Msun /h$ and $7.4 \times 10^6 \Msun/h$, respectively. The force softening scale of the dark matter and stellar particles ($1.48$ kpc$/h$ at $z < 1$) is much smaller than the scales we study in this paper. We choose the simulation snapshot at redshift 0.4, which is a representative redshift of current and future galaxy surveys. The simulations use the following cosmological parameters: $\Omega_{\Lambda}=0.6911$, $\Omega_M=0.3089$, $\Omega_b=0.0486$, $\sigma_8= 0.8159$, and $h=0.6774$.

\subsection{Outer Stellar Mass Catalog}

We use the analysis code \textsc{hydrotools} \citep{diemerCOLOSSUSPythonToolkit2018, diemerAtomicMolecularGas2019} to produce projected stellar mass maps for the galaxies in TNG300-1.  Starting from the center of each galaxy, we select all stellar particles within a 300 kpc cube that belong to the FoF group but do not belong to a satellite galaxy, following \citet{Xu25, Zhou25, leidigReachingEdgeII2025}.  This criterion avoids underestimating the stellar content in the outskirts compared to using \textsc{subfind} gravitational binding alone, since the binding criterion becomes incomplete at large radii.  We then project all selected particles along the $z$-axis to produce $300 \times 300$ kpc$^2$ stellar mass maps with 300 pixels per side.

\begin{figure}[h]
    \centering
    \includegraphics[width=0.8\linewidth]{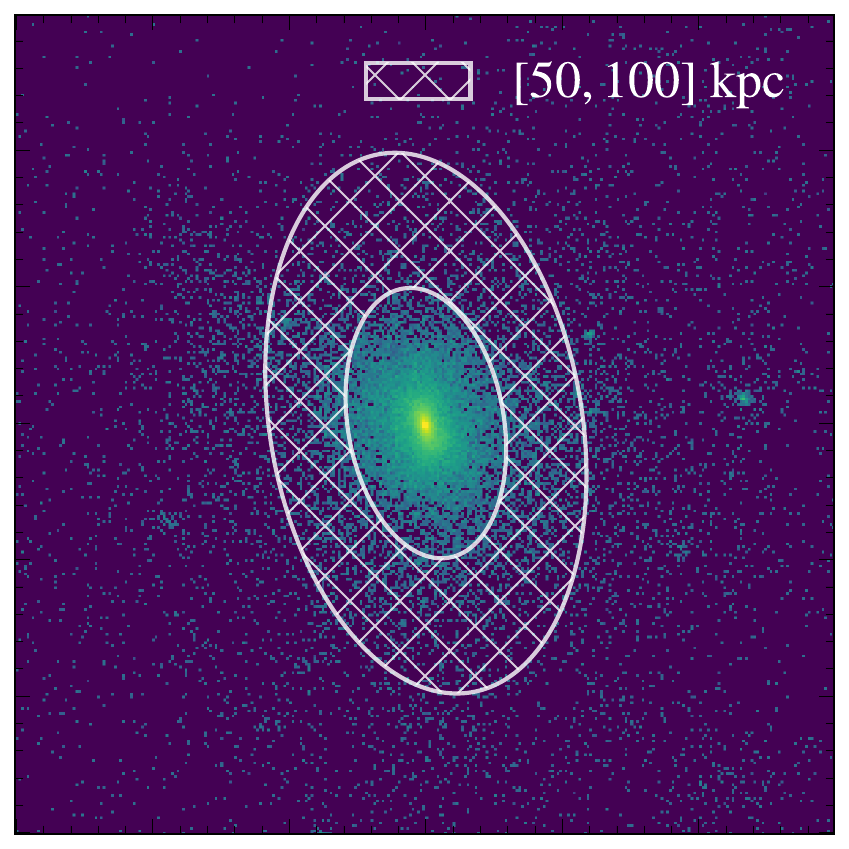}
    \caption{Example of outer stellar mass measurement in the simulation. The white hatched region denotes the outer stellar mass selection: an elliptical annulus with inner and outer semi-major axis lengths of 50 and 100 kpc, whose shape is determined by the overall shape of the galaxy.}
    \label{fig:outermass}
\end{figure}

To measure the 2D outer stellar mass, we follow the isophotal fitting procedure of \citet{Xu25} and \citet{Zhou25}.  We use the method of \citet{jedrzejewskiCCDSurfacePhotometry1987} as implemented in \textsc{photutils} \citep{bradleyPhotutilsPhotometryTools2016} to extract elliptical isophotes from each stellar mass map.  The semi-major axis is grown in steps of 10\% of its current length, with $2\sigma$ clipping applied three times to mask outlier pixels, and the ellipse center is fixed at the projected galaxy center throughout.  From the resulting isophotes, we compute the flux-weighted mean ellipticity and position angle.  These define an elliptical annulus, and we measure the outer stellar mass $M_\mathrm{out}$ as the integrated stellar mass within this annulus between semi-major axis lengths of 50 and 100 kpc.  An example outer mass annulus from the simulations is shown in Figure \ref{fig:outermass}.

\citet{Zhou25} show that this 2D elliptical annulus definition is the optimal 2D selection for minimizing stellar-to-halo mass relation (SHMR) scatter, with a best-case scatter of $\sim 0.2$ dex.  Furthermore, they find that 2D and 3D outer stellar mass selections produce nearly identical galaxy-galaxy lensing profiles, demonstrating that the 2D projection does not introduce major systematic projection effects in lensing.

\section{Percolation Methodology}
\label{sec:method}
We expect satellite contamination to be a dominating systematic effect in using outer stellar mass as a method for cluster selection, especially at lower masses.  However, it is possible to reduce this contamination by filtering based on proximity, and here we explore the effectiveness of a simple percolation method at removing satellite galaxies. Here satellite galaxies are defined as the subhalos of a friends-of-friends (FoF) group other than its central (most massive) subhalo.

First, we start with the simulated catalog of galaxies from TNG300 along with their masses and their projected positions.  The percolation is applied to real-space positions projected along the $z$-axis through the full simulation box; no line-of-sight selection is applied.
Then the percolation program begins by setting an outer mass threshold, which filters the catalog for any galaxy smaller than the selected threshold; we run the program with multiple outer mass thresholds to test the efficacy of percolation as a function of mass. Following mass filtering, we sort the galaxies in the catalog in descending order according to their outer stellar masses.

The first step in the percolation algorithm is to select a target galaxy from the catalog, which begins with the highest outer stellar mass galaxy in our sorted catalog and then works down in mass from there. Next, the program determines a percolation radius; in this work we choose percolation radii based on the $R_{200c}$ radius of the halo to which the target galaxy belongs.  We explore several choices of percolation radii ranging from $0.75 R_{200c}$ to $3.0 R_{200c}$.
Then the percolation algorithm identifies any galaxy, other than the target galaxy, within the percolation radius and removes it from the catalog of galaxies. These steps then repeat as the algorithm iterates through the catalog. Once this is done, the percolation process is complete, and the code saves the now percolated catalog.
Thus, the algorithm removes all galaxies with lower outer stellar mass surrounding the larger outer stellar mass galaxies.

This procedure should remove any satellite galaxies within the percolation radius as long as their outer stellar masses are lower than that of the central galaxy.  We now evaluate the reduction in the satellite fraction as a function of mass threshold and percolation radius.

\section{Results}
\label{sec:results}

In this section, we report results on the satellite fraction and massive halo selection rate as a function of percolation radius and initial outer stellar mass cut for an outer stellar mass cluster selection.

\subsection{Satellite Fraction}

\begin{figure}[h]
    \centering
    \includegraphics[width=1.0\linewidth]{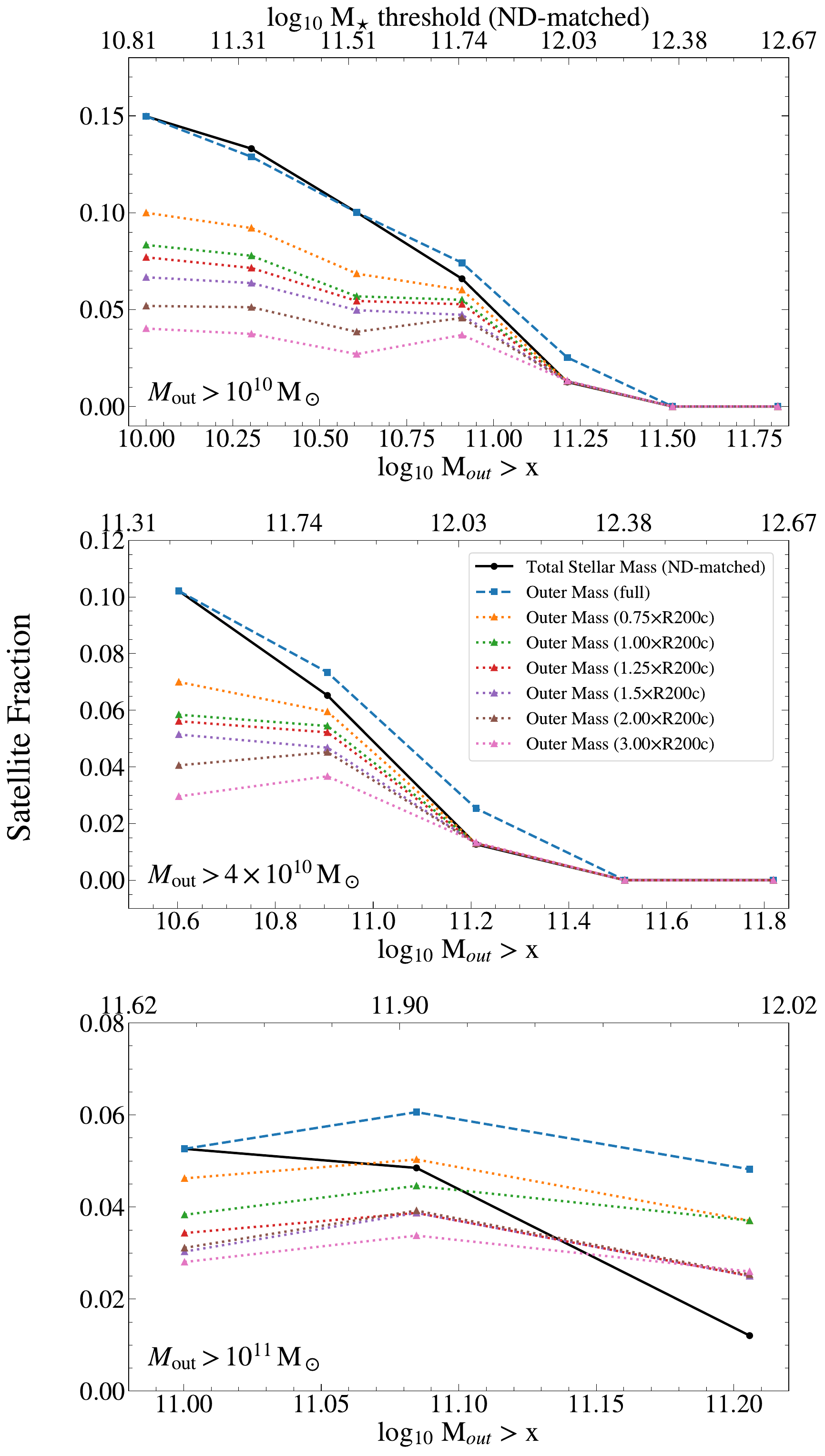}
    \caption{Fraction of galaxies with outer stellar mass greater than the value shown on the x-axis that are satellite galaxies. The dashed blue line shows no percolation, while the colored dotted lines show percolation radii from $0.75 R_{200c}$ to $3.0 R_{200c}$. The solid black line shows a total stellar mass selected sample matched in number density to the outer stellar mass sample. The different panels show the results for different initial outer stellar mass thresholds before percolation: $10^{10} \mathrm{M}_{\odot}$ (top), $4 \times 10^{10} \mathrm{M}_{\odot}$ (middle), and $10^{11} \mathrm{M}_{\odot}$ (bottom).}
    \label{fig:percolation}
\end{figure}

Figure \ref{fig:percolation} shows the satellite fraction, defined as the fraction of galaxies in the sample that are FoF satellites, for an outer stellar mass selected sample after percolation for three different initial thresholds on outer stellar mass of $10^{10} \mathrm{M}_{\odot}$ (top panel), $4 \times 10^{10} \mathrm{M}_{\odot}$ (middle panel), and $10^{11} \mathrm{M}_{\odot}$ (bottom panel). For reference, previous work has found $M_\mathrm{out} = 4 \times 10^{10} \mathrm{M}_{\odot}$ corresponds roughly to $M_\mathrm{vir} \sim 7 \times 10^{13} \mathrm{M}_{\odot}$, and $M_\mathrm{out} = 10^{11} \mathrm{M}_{\odot}$ corresponds roughly to $M_\mathrm{vir} \sim 2 \times 10^{14} \mathrm{M}_{\odot}$ \citep{kwiecienImprovingGalaxyCluster2025}. The satellite fraction is plotted as a function of minimum outer stellar mass for no percolation and percolation radii ranging from $0.75 R_{200c}$ to $3.0 R_{200c}$.  In general, the satellite fraction is low even for no percolation; it is always $\leq$15\% even down to $M_\mathrm{out} > 10^{10} \mathrm{M}_{\odot}$, and for $M_\mathrm{out} > 10^{11} \mathrm{M}_{\odot}$ it is less than 10\%.  Adding percolation successfully reduces the satellite fraction, and the removal of satellites improves with a larger percolation radius all the way to $R_\mathrm{perc} = 3.0 R_{200c}$.  For the largest percolation radii, the satellite fraction is reduced to $<$4\% for the sample with $M_\mathrm{out} > 10^{11} \mathrm{M}_{\odot}$.

Figure \ref{fig:percolation} also shows the satellite fraction for a total stellar mass selected sample with the same number density as the corresponding outer mass sample.  Without percolation, the satellite fractions are similar for total and outer stellar mass; with percolation, our samples perform significantly better at lower masses, while at the highest masses the satellite fractions of both selections are small.

\subsection{Massive Halo Selection}

\begin{figure}[h]
    \centering
    \includegraphics[width=1.0\linewidth]{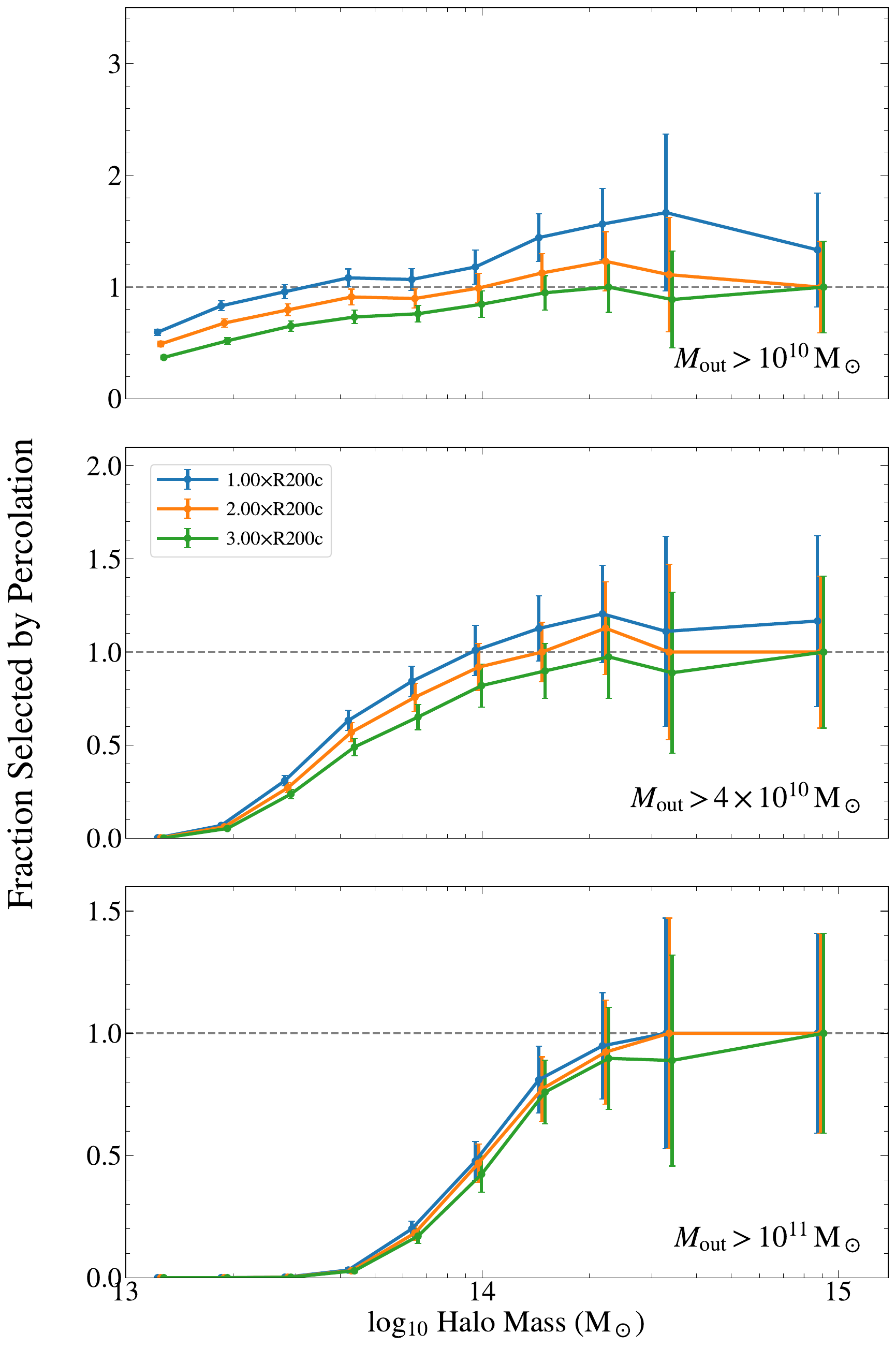}
    \caption{Ratio of the number of galaxies (centrals+satellites) selected by percolation to the number of central galaxies in bins of halo mass for percolation radii of $1.0 R_{200c}$ (blue), $2.0 R_{200c}$ (yellow), and $3.0 R_{200c}$ (green). \textit{Top:} Initial outer stellar mass cut of $10^{10} \mathrm{M}_{\odot}$. \textit{Middle:} Initial outer stellar mass cut of $4 \times 10^{10} \mathrm{M}_{\odot}$. \textit{Bottom:} Initial outer stellar mass cut of $10^{11} \mathrm{M}_{\odot}$.}
    \label{fig:allhalos}
\end{figure}

We now ask which halos are selected by outer stellar mass after percolation.  Ideally, our cluster selection would select a complete sample of all halos above some mass threshold without contamination. Over-percolation or an overly aggressive outer stellar mass cut may result in us losing massive halos from the sample, while remaining satellites that are not removed will result in double-counting halos.  Figure \ref{fig:allhalos} shows the fraction of halos in bins of $M_{200c}$ halo mass selected after percolation for percolation radii of 1.0, 2.0, and 3.0 $R_{200c}$ and for initial thresholds on outer stellar mass of $10^{10} \mathrm{M}_{\odot}$ (top panel), $4 \times 10^{10} \mathrm{M}_{\odot}$ (middle panel), and $10^{11} \mathrm{M}_{\odot}$ (bottom panel). Here the fraction might be larger than one if both a central galaxy and a satellite galaxy belonging to the same halo remain in the catalog after percolation.  Error bars reflect counting statistics on the ratio given the number of percolated galaxies and halos in a particular bin.

For the two lower outer stellar mass cuts and a percolation radius of $1.0 R_{200c}$, there is some duplication of the most massive halos due to remaining satellite galaxies in the sample that preferentially live in the most massive halos.  This contamination is mostly eliminated by using a larger percolation radius. 
Using a higher outer stellar mass threshold also significantly reduces the satellite contamination.  An $M_\mathrm{out} > 10^{11} \mathrm{M}_{\odot}$ cut gives a sample without duplication for all percolation radii. 
However, this cut limits us to higher mass clusters.  In all three cases, incompleteness sets in at lower halo masses, but for the lower outer mass thresholds and percolation radii up to $2.0 R_{200c}$ the fractions only dip below 100\% for halo masses less than $10^{14} \mathrm{M}_{\odot}$, while for $M_\mathrm{out} > 10^{11} \mathrm{M}_{\odot}$, this occurs at $M_{200c} < 3\times10^{14} \mathrm{M}_{\odot}$.  It should be noted that the number of massive halos in the simulations is small, and the uncertainties due to counting statistics on the recovery fractions are, therefore, large.

\begin{figure}[h]
    \centering
    \vspace{11pt}
    \includegraphics[width=1.0\linewidth]{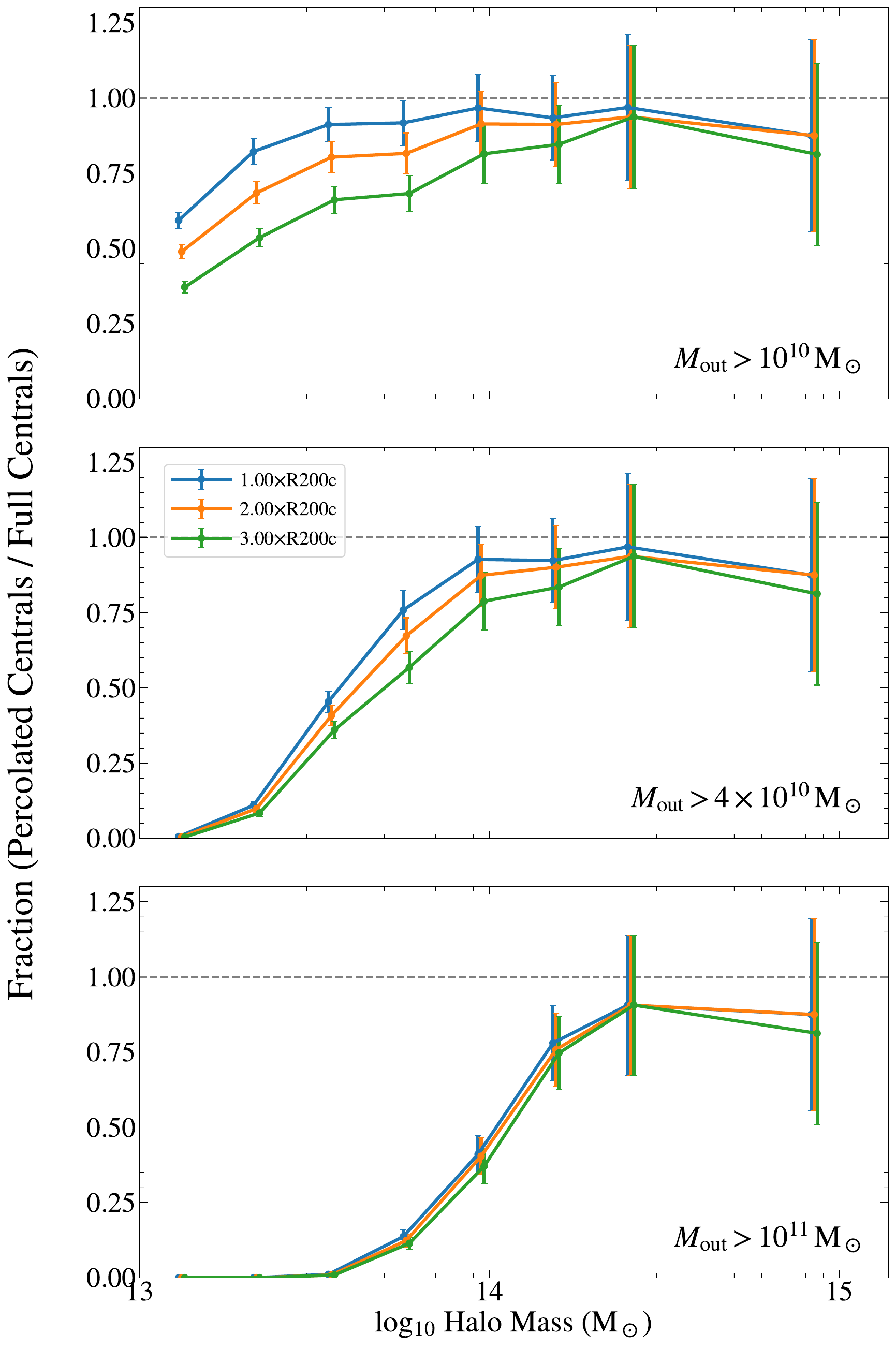}
    \caption{Fraction of central galaxies selected by percolation in bins of halo mass for percolation radii of $1.0 R_{200c}$ (blue), $2.0 R_{200c}$ (yellow), and $3.0 R_{200c}$ (green). \textit{Top:} Initial outer stellar mass cut of $10^{10} \mathrm{M}_{\odot}$. \textit{Middle:} Initial outer stellar mass cut of $4 \times 10^{10} \mathrm{M}_{\odot}$. \textit{Bottom:} Initial outer stellar mass cut of $10^{11} \mathrm{M}_{\odot}$.}
    \label{fig:centralhalos}
\end{figure}

Halos may appear in the sample because their central is selected or because a satellite in that halo is selected. In Figure \ref{fig:centralhalos}, we look at the completeness for central galaxies after percolation. Here we see that the recovery fractions are less than 100\%, meaning that we are losing some central galaxies in the percolation process.  However, for the most massive halos the completeness is $\gtrsim 80$\% for all initial mass cuts and percolation radii.  
Again, small-number statistics in the more massive bins mean that the uncertainties on the completeness are relatively large.

\begin{figure}[h]
    \centering
    \includegraphics[width=1.0\linewidth]{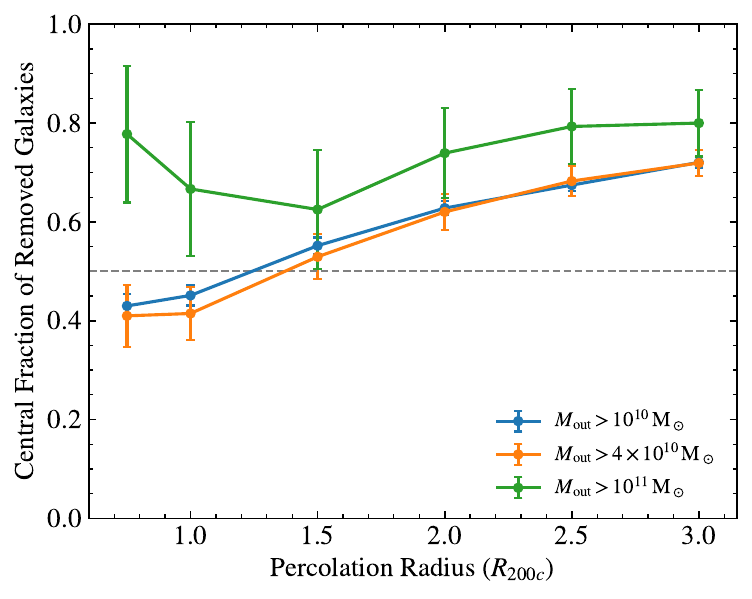}
    \caption{Fraction of the galaxies removed by percolation that are central galaxies as a function of percolation radius for initial outer stellar mass cuts of $10^{10} \mathrm{M}_{\odot}$ (blue), $4 \times 10^{10} \mathrm{M}_{\odot}$ (orange), and $10^{11} \mathrm{M}_{\odot}$ (green).  The horizontal dashed line marks 50\%, where the removed population transitions from satellite-dominated to central-dominated.  Error bars reflect binomial counting statistics given the number of removed galaxies.  The removed centrals are predominantly centrals of group-scale halos ($M_{200c} < 10^{14} \mathrm{M}_{\odot}$) or centrals removed through chance line-of-sight alignments, so the completeness for cluster-scale halos remains high (see text).}
    \label{fig:removedfrac}
\end{figure}

Figure \ref{fig:removedfrac} shows the composition of the galaxies in the removed population: the fraction of the galaxies removed by percolation that are central galaxies, as a function of percolation radius for the three initial outer stellar mass cuts.  For the $10^{10} \mathrm{M}_{\odot}$ and $4 \times 10^{10} \mathrm{M}_{\odot}$ cuts, the removed population is satellite-dominated for percolation radii below $\approx 1.3 R_{200c}$ and central-dominated above it, with centrals making up $\approx 60$--$72\%$ of the removed galaxies for radii of $2.0$--$3.0 R_{200c}$.  For the $10^{11} \mathrm{M}_{\odot}$ cut, the removed population is central-dominated at all radii, although the total number of removed galaxies is small ($9$--$35$).

The centrals removed at large radii are predominantly centrals of group-scale halos that lie near a more massive galaxy in projection: for the $4 \times 10^{10} \mathrm{M}_{\odot}$ cut and a percolation radius of $2.0 R_{200c}$, the removed centrals have a median halo mass of $\sim 4.5 \times 10^{13} \mathrm{M}_{\odot}$, $\approx 88\%$ of them have $M_{200c} < 10^{14} \mathrm{M}_{\odot}$, and only 13 of the 175 cluster-scale ($M_{200c} > 10^{14} \mathrm{M}_{\odot}$) centrals in the sample are removed, consistent with the high completeness for massive halos in Figure \ref{fig:centralhalos}.  Of these 13, seven are removed by a satellite of their own halo with higher outer stellar mass, so the halo remains in the sample represented by that satellite, though it would be miscentered. The remaining six are removed by galaxies separated by $4$--$100$ Mpc along the line of sight---chance alignments that arise because the percolation is applied in projection through the full simulation box with no line-of-sight selection.  In survey data, redshift information would exclude the most distant of these alignments.  For this cut and percolation radius, percolation therefore removes few cluster-scale halos from the sample, as can be seen in Figures \ref{fig:allhalos} and \ref{fig:centralhalos}.  We note, however, that the situation in real data is more complicated: peculiar velocities and redshift uncertainties limit how finely galaxies can be separated along the line of sight, and we leave a redshift-space implementation of the percolation to future work.  The incompleteness could also be improved by cross-matching with other cluster catalogs.

\section{Discussion}
\label{sec:discussion}

Our results indicate that overall satellite contamination in outer stellar mass selected samples is low and is considerably improved by proximity-based percolation, but some satellite contamination remains for the most massive halos, which can host more than one large outer stellar mass galaxy. However, it can be seen from Figure \ref{fig:percolation} that the TNG300 simulations used here include satellite galaxies out to very large radii $>3.0 R_{200c}$.  This could be a result of the FoF halo finding employed, which can link well-separated halos and galaxies into the same parent halo.  

\begin{figure*}[t]
    \centering
    \includegraphics[width=1.0\linewidth]{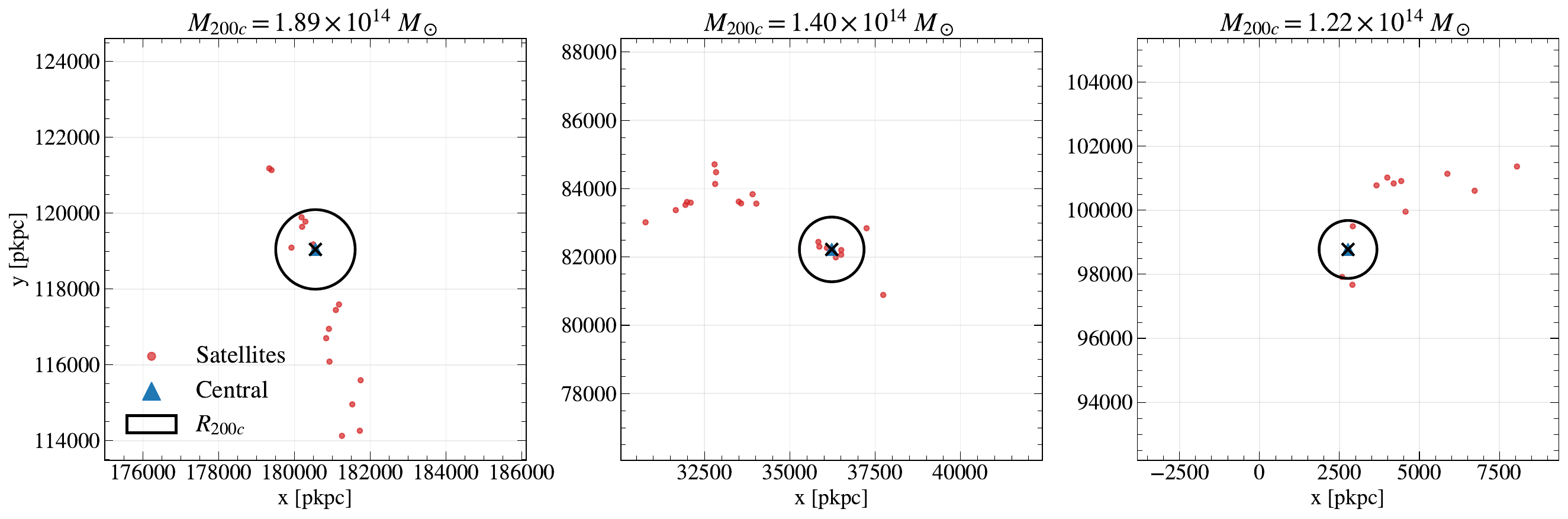}
    \caption{Projected distribution of the friends-of-friends (FoF) member galaxies of three example cluster-scale halos ($M_{200c}\approx 1.2$--$1.9\times10^{14}\,\mathrm{M}_{\odot}$, labeled and ordered by descending mass) in TNG300 at $z=0.4$. Only galaxies belonging to the same FoF group as the central are shown: the central is marked by the blue triangle with a black cross at the halo center, satellites are shown in red, and the black circle indicates $R_{200c}$. Each panel is scaled independently to its own satellite extent. These groups are selected to illustrate FoF over-linking: many of their member satellites lie far outside $R_{200c}$, reaching $\approx 5$--$6.5\,R_{200c}$ in projection. Such galaxies are bound to the same nominal halo by the FoF finder but would not observationally be associated with the cluster, so the satellite contamination measured in this work is likely overestimated.}
    \label{fig:fof}
\end{figure*}

To quantify this, we assign FoF membership using the group $M_{200c}$, which is shared by all galaxies belonging to the same FoF group, and measure the projected separation of each satellite from its central in units of the halo $R_{200c}$.  Figure \ref{fig:fof} shows three cluster-scale halos ($M_{200c}\approx 1.2$--$1.9\times10^{14}\,\mathrm{M}_{\odot}$) chosen to have spatially extended FoF membership; among FoF groups with at least ten satellites, we select those with the largest number of members beyond $2\,R_{200c}$ in projection. Many of these FoF member galaxies lie well outside $R_{200c}$, reaching $\approx 5$--$6.5\,R_{200c}$ in projection, with $8$--$13$ satellites per group beyond $2\,R_{200c}$; these distant members are associated with the central halo only through FoF linking.

Such extended membership is not limited to these examples.  Across the catalog, $\approx 40\%$ of satellite galaxies lie beyond $R_{200c}$ of their host center, $\approx 18\%$ beyond $2\,R_{200c}$, and $\approx 5\%$ beyond $3\,R_{200c}$, with a tail extending to much larger separations.

Observationally, it is unlikely that these distant galaxies would be associated with the same cluster.  The percolation removes galaxies by projected proximity to a more massive galaxy, and these distant FoF members can survive percolation even for large percolation radii while genuine close satellites are successfully removed.  We therefore expect that the satellite contamination found here could be overestimated compared to an observed cluster catalog.

\section{Conclusions}
\label{sec:conclusions}
The presence of an extended stellar halo in massive galaxies, specifically the stellar mass in an annulus from $50$--$100$ kpc, has emerged as a promising method of selecting clusters of galaxies that is both relatively free of selection effects, like the projection effects which plague red-sequence selection methods, and provides a low-scatter cluster mass proxy \citep{huangOuterStellarMass2021, kwiecienImprovingGalaxyCluster2025, Zhou25}.  Using the IllustrisTNG300 simulation and treating non-central FoF subhalos as satellites, we test the prevalence of satellite contamination, the primary systematic for this selection technique, and the efficacy of reducing satellite contamination with a simple percolation method.

Even without percolation, the satellite fraction is low, $\leq$15\% for $M_\mathrm{out} > 10^{10} \mathrm{M}_{\odot}$, and $\leq$6\% for $M_\mathrm{out} > 10^{11} \mathrm{M}_{\odot}$. The satellite contamination can be effectively reduced by a simple, proximity-based percolation that removes lower outer stellar mass galaxies within some radius of a higher-mass galaxy.  We test percolation radii ranging from $0.75 R_{200c}$ to $3.0 R_{200c}$.  The satellite fraction consistently decreases with increasing percolation radius.  The presence of massive satellite galaxies at very large radii ($>3.0 R_{200c}$) from the center of the halo is likely an artifact of the FoF halo finding used in TNG300 and means that our satellite fractions after percolation are likely overestimated.

The satellite contamination can also be effectively reduced by employing a higher mass cutoff on the outer stellar mass selection at the cost of increasing the mass at which the halo selection has high completeness.

Overall, we find that for a moderately high outer stellar mass cut ($\sim 4 \times 10^{10} \mathrm{M}_{\odot}$) and a percolation radius of $\sim2.0 R_{200c}$, we can select a cluster catalog with both high completeness and purity for halo masses $\gtrsim 10^{14} \mathrm{M}_{\odot}$.  Therefore, outer stellar mass shows promise as a cluster selection technique.

Applying the percolation method presented here in real data will require utilizing an observationally determined percolation radius.  As outer stellar mass is a low-scatter proxy for halo mass, it is possible to estimate an $R_{200c}$ radius based on the observed outer stellar mass.  While this estimated radius will have uncertainty, Figures \ref{fig:allhalos} and \ref{fig:centralhalos} show that near the adopted radius of $\sim 2.0 R_{200c}$ the purity and completeness vary slowly with the exact radius used.  We, therefore, expect this method to be effective in observational survey data, and it can be validated using multiwavelength data like X-ray and Sunyaev-Zeldovich effect observations.

\section*{Software Acknowledgements}
This work made use of the following software packages: \texttt{numpy} \citep{harrisArrayProgrammingNumPy2020}, \texttt{scipy} \citep{virtanenSciPy10Fundamental2020}, \texttt{pandas} \citep{mckinneyDataStructuresStatistical2010}, \texttt{matplotlib} \citep{hunterMatplotlib2DGraphics2007}, \texttt{h5py} \citep{collettePythonHDF52013}, \texttt{astropy} \citep{theastropycollaborationAstropyProjectBuilding2018}, \texttt{photutils} \citep{bradleyPhotutilsPhotometryTools2016}, \texttt{colossus} \citep{diemerCOLOSSUSPythonToolkit2018}, and \texttt{jupyter} \citep{grangerJupyterThinkingStorytelling2021}.
\bibliographystyle{mnras}
\bibliography{betterbibtex,zhou}

\end{document}